%% file: bleadv.tex
\documentclass[sigconf,authorversion]{acmart}

\usepackage{placeins} 

\usepackage{adjustbox}

\usepackage{stfloats}

\usepackage[colorinlistoftodos,prependcaption]{todonotes} 
\presetkeys%
    {todonotes}%
    {inline,backgroundcolor=orange}{}
    
\usepackage{fontawesome}
\usepackage[nopostdot,acronym,shortcuts,nonumberlist]{glossaries}
\newacronym{(S+N)/NR}{(S+N)/NR}{signal-plus-noise-to-noise ratio}
\newacronym{ADC}{ADC}{analog-to-digital converter}
\newacronym{BCI}{BCI}{bulk current injection}
\newacronym{CAN}{CAN}{control area network}
\newacronym{CRT}{CRT}{cathode ray tube}
\newacronym{CSMACD}{CSMA/CD}{carrier sense multiple access with collision detection}
\newacronym{CoE}{CoE}{CAN over Ethernet}
\newacronym{DAC}{DAC}{digital-to-analog converter}
\newacronym{DoS}{DoS}{Denial of Service}
\newacronym{EMR}{EMR}{electromagnetic radiation}
\newacronym{FCS}{FCS}{frame check sum}
\newacronym{FEC}{FEC}{forward error correction}
\newacronym{FPGA}{FPGA}{field-programmable gate array}
\newacronym{ICMP}{ICMP}{Internet control message protocol}
\newacronym{IEC}{IEC}{International Electrotechnical Commission}
\newacronym{MAC}{MAC}{Medium Access Control}
\newacronym{NIC}{NIC}{network interface card}
\newacronym{NSA}{NSA}{National Security Agency}
\newacronym{PA}{PA}{power amplifier}
\newacronym{PC}{PC}{personal computer}
\newacronym{RF}{RF}{radio frequency}
\newacronym{SDR}{SDR}{software-defined radio}
\newacronym{SFD}{SFD}{start frame delimiter}
\newacronym{SNR}{SNR}{signal-to-noise ratio}
\newacronym{USRP}{USRP}{Universal Software Radio Peripheral}
\newacronym{SSP}{SSP}{Secure Simple Pairing}
\newacronym{USB}{USB}{Universal Serial Bus}
\newacronym{TLV}{TLV}{Type Length Value}
\newacronym{L2CAP}{L2CAP}{Logical Link Control and Adaptation Protocol}
\newacronym{AOP}{AOP}{Always-On Processor}
\newacronym{ACL}{ACL}{Asynchronous Connec\-tion-Less}
\newacronym{BLE}{BLE}{Bluetooth Low Energy}
\newacronym{BR}{BR}{Basic Rate}
\newacronym{CID}{CID}{Channel ID}
\newacronym{PDU}{PDU}{Protocol Data Unit}
\newacronym{B-Frame}{B-Frame}{Basic Information Frame}
\newacronym{C-Frame}{C-Frame}{Control Frame}
\newacronym{PSM}{PSM}{Protocol/Service Multiplexer}
\newacronym{HCI}{HCI}{Host Controller Interface}
\newacronym{EDR}{EDR}{Enhanced Data Rate}
\newacronym{SCO}{SCO}{Synchronous Connection-Oriented}
\newacronym{HSP}{HSP}{Headset Profile}
\newacronym{ATT}{ATT}{Attribute Protocol}
\newacronym{TCP}{TCP}{Transmission Control Protocol}
\newacronym{SoC}{SoC}{System-on-a-Chip}
\newacronym{IoT}{IoT}{Internet of Things}
\newacronym{GPS}{GPS}{Global Positioning System}
\newacronym{CERT}{CERT}{Computer Emergency Response Team}
\newacronym{API}{API}{Application Programming Interface}
\newacronym{MITM}{MITM}{Machine-in-the-Middle}
\newacronym{TLS}{TLS}{Transport Layer Security}
\newacronym{OTA}{OTA}{Over-the-Air}
\newacronym{XSS}{XSS}{Cross-Site Scripting}
\newacronym{SDK}{SDK}{Software Development Kit}
\newacronym{MQTT}{MQTT}{Message Queuing Telemetry Transport}
\newacronym{CVE}{CVE}{Common Vulnerabilities and Exposures}
\newacronym{ECDH}{ECDH}{Elliptic-curve Diffie–Hellman}
\newacronym{AEAD}{AEAD}{Authenticated Encryption with Associated Data}
\newacronym{AES}{AES}{Advanced Encryption Standard}
\newacronym{TOR}{TOR}{The Onion Router}
\newacronym{GDPR}{GDPR}{EU General Data Protection Regulation}
\newacronym{HSTS}{HSTS}{HTTP Strict Transport Security}
\newacronym{AWS}{AWS}{Amazon Web Services}
\newacronym{LTK}{LTK}{Long Term Key}
\newacronym{LK}{LK}{Link Key}
\newacronym{RCE}{RCE}{Remote Code Execution}
\newacronym{SIV}{SIV}{Synthetic Initialization Vector}
\newacronym{PoC}{PoC}{Proof of Concept}
\newacronym{XPC}{XPC}{Cross-Process Communication}
\newacronym{MFi}{MFi}{Made for iPhone/iPad/iPod}
\newacronym{ACI}{ACI}{Apple Controller Interface}
\newacronym{ECB}{ECB}{Electronic Codebook}
\newacronym{UART}{UART}{Universal Asynchronous Receiver-Transmitter}
\newacronym{RSSI}{RSSI}{Received Signal Strength Indicator}
\newacronym{BCS}{BCS}{Bluetooth Core Scheduler}

\newacronym{KDF}{KDF}{Key Derivation Function}
\newacronym{PRF}{PRF}{Pseudo-Random Function}

\tikzstyle{line} = [draw, -latex']
\usetikzlibrary{shapes,arrows,calc,positioning,matrix}
\usetikzlibrary{positioning}
\tikzset{
    >=triangle 45
}

\definecolor{darkred}{rgb}{0.831, 0, 0.063}

\usepackage[binary-units]{siunitx}
\usepackage{tabularx}

\definecolor{sorange}{rgb}{0.95, 0.57, 0}
\definecolor{sblue}{RGB}{0, 105, 180}
\colorlet{orange}{sorange}
\colorlet{blue}{sblue}

\usepackage{listings}
\lstdefinelanguage{ASM}{
    morekeywords={b, ble, blt, bne, bx, bl, ldr, str, push, pop, mov, add, sub},
    keywordstyle=\color{blue},
    sensitive=false, 
    morecomment=[l]{//}, 
    morecomment=[s]{/*}{*/}, 
    morestring=[b]", 
} %
\lstdefinelanguage{none}{
  identifierstyle=
}

\usepackage{subcaption}

\usepackage{xspace}

\usepackage{bytefield}
\usepackage{xcolor,colortbl}
	
\acmYear{2020}\copyrightyear{2020}
\setcopyright{rightsretained}
\acmConference[WiSec '20]{13th ACM Conference on Security and Privacy in Wireless and Mobile Networks}{July 8--10, 2020}{Linz (Virtual Event), Austria}
\acmBooktitle{13th ACM Conference on Security and Privacy in Wireless and Mobile Networks (WiSec '20), July 8--10, 2020, Linz (Virtual Event), Austria}
\acmPrice{15.00}
\acmDOI{10.1145/3395351.3401699}
\acmISBN{978-1-4503-8006-5/20/07}

\acmSubmissionID{4}

\begin{document}

\title{DEMO: Extracting Physical-Layer BLE Advertisement Information from Broadcom and Cypress Chips}


\author{Jiska Classen}

\affiliation{%
  \institution{Secure Mobile Networking Lab}
  \institution{TU Darmstadt, Germany}
  \streetaddress{Pankratiusstraße 2}
  \postcode{64289}
}
\email{jclassen@seemoo.de}

\author{Matthias Hollick}

\affiliation{%
  \institution{Secure Mobile Networking Lab}
  \institution{TU Darmstadt, Germany}
  \streetaddress{Pankratiusstraße 2}
  \postcode{64289}
}
\email{mhollick@seemoo.de}

\renewcommand{\shortauthors}{Classen and Hollick}

\input{chapters/abstract.tex}



\maketitle

\input{chapters/intro.tex}
\input{chapters/reversing.tex}
\input{chapters/conclusion.tex}
\input{chapters/demo.tex}

\begin{acks}
We thank Alexander Heinrich for pointing out the issue that all advertisements are reported to be received on channel 37.

This work has been funded by the German Federal Ministry of Education and Research and the Hessen State Ministry for Higher Education, Research and the Arts within their joint support of the National Research Center for Applied Cybersecurity ATHENE.
\end{acks}

\bibliographystyle{ACM-Reference-Format}
\bibliography{bibfile}

\end{document}

%% file: chapters/abstract.tex

\begin{abstract}

Multiple initiatives propose utilizing \ac{BLE} advertisements for contact tracing and SARS-CoV-2 exposure notifications. This demo shows a research tool to analyze \ac{BLE} advertisements; if universally enabled by the vendors, the uncovered features could improve exposure notifications for everyone. We reverse-engineer the firmware-internal implementation of \ac{BLE} advertisements on \emph{Broadcom} and \emph{Cypress} chips and show how to extract further physical-layer information at the receiver. The analyzed firmware works on hundreds of millions of devices, such as all \emph{iPhones}, the European \emph{Samsung Galaxy S} series, and \emph{Raspberry Pis}.
\end{abstract}

%% file: chapters/intro.tex

\vspace{2em} 
\section{Introduction}

Specification-compliant \ac{BLE} advertisements contain little information that can be used for proximity estimation.
We assume that the \emph{Apple} and \emph{Google} exposure notification \ac{API} uses advertisements nonetheless~\cite{applegooglebt},
because they preserve battery and limit \ac{RCE} risks.
When receiving an advertisement, the chip measures the \ac{RSSI} and includes it in the advertisement report
before forwarding it to the operating system.
This behavior is specification-compliant and does not need any further
modifications~\cite[p. 2382]{bt52}.
There are multiple ways to enhance proximity measurements. For example, one could include the sender’s transmission power in the advertisements~\cite[p. 4]{applegooglebt}.
The transmission power directly influences
the \ac{RSSI} on the receiver. Not all chips can be set to the same transmission power, and thus, \ac{RSSI} measurements need to be adjusted 
on the receiver.

\begin{figure}[!b]
\center

	\begin{center}
	\begin{tikzpicture}[minimum height=0.55cm, scale=0.8, every node/.style={scale=0.8}, node distance=0.7cm]

    \node[inner sep=0pt] (iphone) at (-2,0)
    {\includegraphics[height=1.5cm]{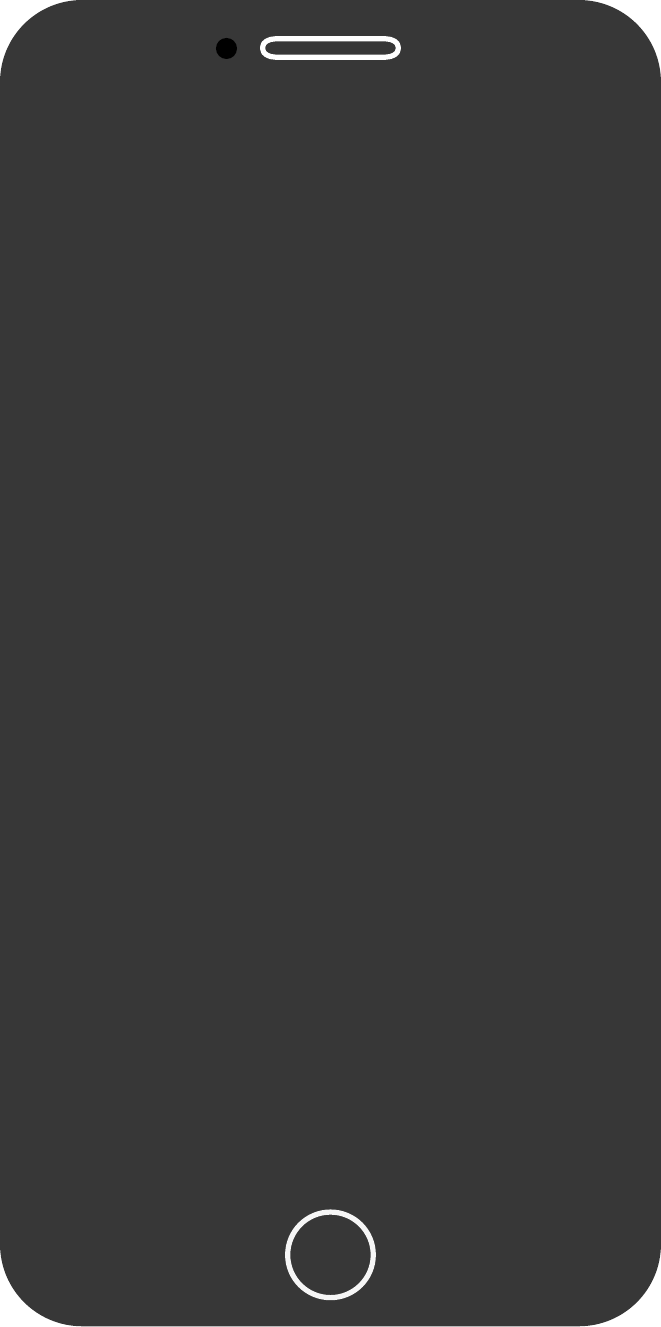}}; 
    \node[inner sep=0pt] (iphonex) at (-2,0)
    {\includegraphics[height=1.2cm]{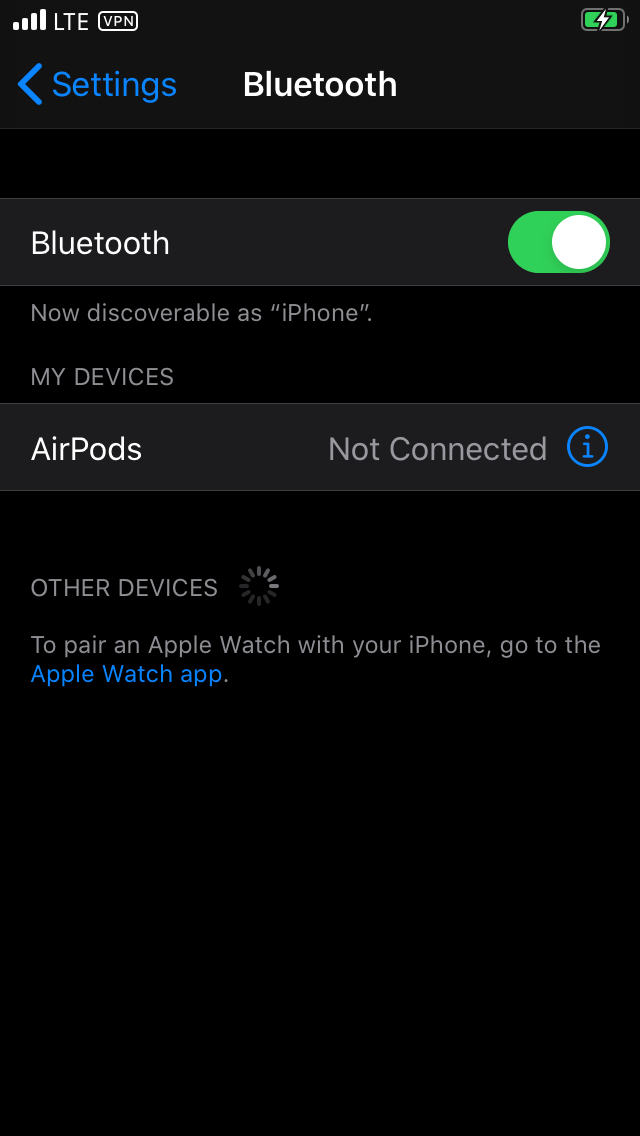}};  
    \node[inner sep=0pt] (chipa) at (-1,-0.55)
    {\includegraphics[height=0.4cm]{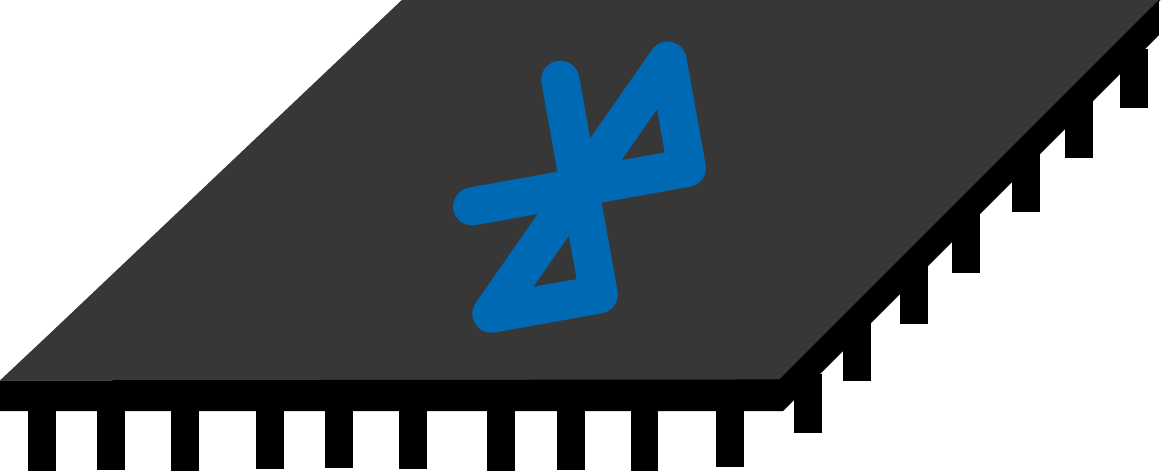}};    
   	
    \node[inner sep=0pt] (macos) at (6,0.5)
    {\includegraphics[height=1.5cm]{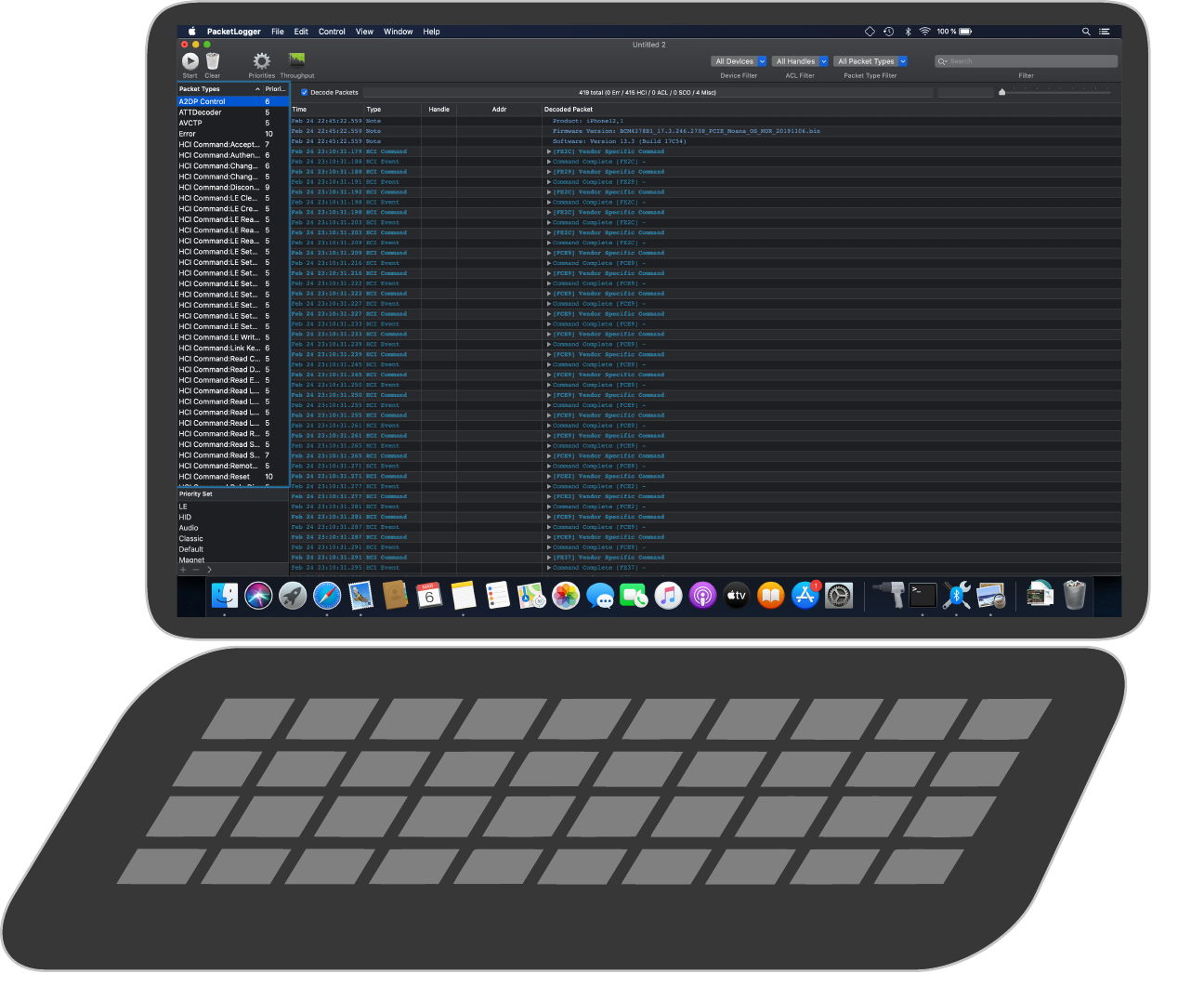}}; 
    \node[inner sep=0pt] (ib) at (5.8,-0.55)
    {\includegraphics[height=0.4cm]{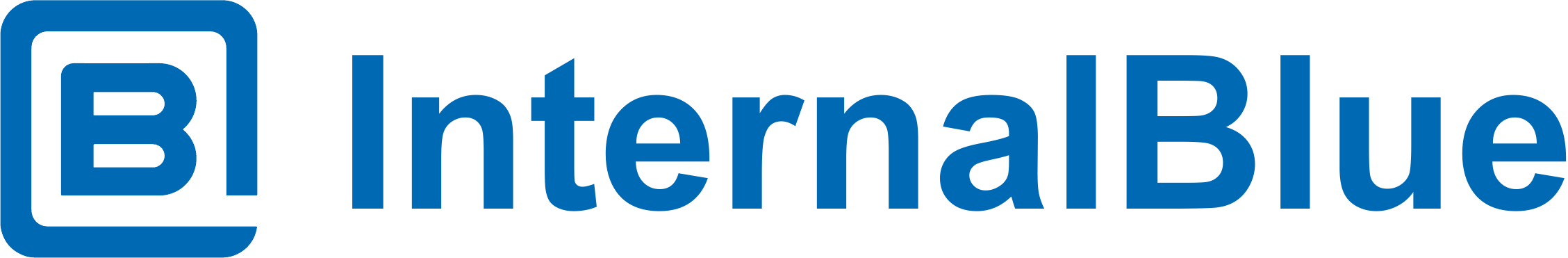}}; 
    \node[inner sep=0pt] (chipa) at (4,-0.55)
    {\includegraphics[height=0.4cm]{pics/bt_wifi_chip.pdf}};

    \path[->] (-1.5,0.3) edge node[sloped, anchor=center, above, text width=3cm,xshift=-0.3cm] {BLE Advertisement \\ \textcolor{gray}{Optional: TX Power}} (5.2,0.3);
    
    \path[->,color=blue] (-0.4,-0.5) edge node[sloped, anchor=center, above, text width=3cm,yshift=-0em] {\textcolor{blue}{PHY Channel: 37..39}} (3.5,-0.5);

	\end{tikzpicture}
	\end{center}
	
\caption{\ac{BLE} advertisement physical-layer information.}
\label{fig:bleadv}
\end{figure}

We find that \emph{Apple's}
Bluetooth \emph{PacketLogger} goes beyond the Bluetooth specification and displays the channel, antenna, and scan mode of each advertisement---but this information is always set to zero.
Further investigation reveals that this information can be enabled with a firmware-internal global variable.
As shown in \autoref{fig:bleadv}, the advertisements are not modified, as the information is measured on the receiver.
In the most recent version of \emph{InternalBlue}~\cite{mantz2019internalblue}, this
feature can be activated with the \texttt{adv} command.

Information about the advertisement channel is valuable, as the \ac{RSSI} varies depending on the channel.
The three advertisement channels are spread across the spectrum. Thus, typical interference sources such as a Wi-Fi access point can easily be evaded by blacklisting the interfered channel without taking statistics on multiple advertisements over time.

This demo is structured as follows. We explain the detailed reverse-engineering workflow to 
uncover proprietary features in \autoref{sec:reverse}. In \autoref{sec:conclusion}, we conclude our findings.

%% file: chapters/reversing.tex

\begin{table*}[!t]
		\footnotesize
     	\fontfamily{\ttdefault}\selectfont
     	\center

	    \caption{\ac{BLE} advertisement report captured with \emph{PacketLogger}, including the enhanced event type.}
	    \label{tab:packetlogger}
		\begin{tabular}{ |l|l|l|p{10.7cm}| } 
        \hline
	    \rowcolor{blue!10}
        Apr 15 17:19:28:281 & HCI Event & \textcolor{black!70!black}{CA:FE:BA:BE:13:37} & \textcolor{gray}{$\blacktriangledown$} LE - Advertising Report - 1 Report - Normal - Random - CA:FE:BA:BE:13:37  \newline \hspace*{0.7em} -56 dBm - Channel 37 \\ 
        \hline
        & & & \hspace*{1.2em} Parameter Length: 30 (0x1E) \\
        & & & \hspace*{1.2em} Num Reports: 0x01 \\        
        & & & \hspace*{1.2em} Report 0 \\
        & & & \hspace*{2em} \textcolor{blue!70!black}{Event Type}: \textcolor{blue!70!black}{Scan Mode: Normal Scan Mode} - \textcolor{blue!70!black}{Channel 37}  - \textcolor{blue!70!black}{Antenna: BT}  \newline \hspace*{8em} - Connectable Undirected Advertising  (ADV\_IND) \\        
        & & & \hspace*{2em} \textcolor{gray}{...} \\        
        & & & \hspace*{2em} RSSI: -56 dBm \\
        \hline
	    \rowcolor{blue!10}
        Apr 15 17:19:28:281  & HCI Event & ~ & \textcolor{gray}{$\blacktriangleright$} 00000000: 3E1E 0201 \textcolor{blue!70!black}{00}01 \textcolor{black!70!black}{3713 BEBA FECA} ... \\
        \hline
	    \end{tabular}
\end{table*}

\section{Reverse-Engineering Proprietary Advertisement Features}
\label{sec:reverse}

We use two methods to reverse-engineer vendor-specific additions to the \ac{BLE} advertisement handler.
First, we analyze which information the \emph{PacketLogger}
is using to display the channel in \autoref{ssec:packetlogger}. Based on this information, we analyze the \emph{Broadcom} and \emph{Cypress} firmware
to enable the output of this information in \autoref{ssec:firmware}.

\subsection{PacketLogger}
\label{ssec:packetlogger}
The \emph{PacketLogger} is included in the \emph{Additional Tools for Xcode} on \emph{macOS}.
It features similar functions as \emph{Wireshark} but is specifically designed for Bluetooth in the \emph{Apple} ecosystem.
Thus, it supports various proprietary protocols and features that are a helpful starting point for reverse-engineering.
These protocols are otherwise undocumented. We assume that \emph{Apple} uses the \emph{PacketLogger} for developing
protocols and debugging and, thus, intentionally includes information about these protocols in their toolchain.

\autoref{tab:packetlogger} shows a \ac{BLE} advertisement as captured on \emph{macOS Catalina} with \emph{PacketLogger}.
Note that by default all advertisements are displayed to be on channel 37, even though they are also received on channels 38 and 39.
However, this indicates that there are some means of channel interpretation that are not included in specification-compliant
advertisement reports~\cite[p. 2382]{bt52}.

The \emph{PacketLogger} binary is located in \path{PacketLogger.app/Contents/Frameworks/PacketDecoder.framework/Versions/A/PacketDecoder}.
It contains all the strings displayed within the \emph{PacketLogger} and also most function names, which enables an analysis with \emph{IDA Pro 7.2}.
A search for the string \path{`Channel'}, as it can be seen in the \emph{PacketLogger} output, leads to the function \path{leAdvertisingEventTypeString}.
This function prints the antenna, channel, and scan mode, which are encoded into the upper half byte of
the event type as shown in \autoref{lst:event}.
This is possible because the event type is \SI{1}{\byte}, but Bluetooth specification only defines the values \textcolor{blue!70!black}{\texttt{0x00}--\texttt{0x04}}~\cite[p. 2383]{bt52}.
The channel values \textcolor{blue!70!black}{\texttt{0x0}--\texttt{0x2}} correspond to the Bluetooth channels 37--39. Thus, without this feature enabled on the chip,
the channel is always interpreted as 37 by \emph{PacketLogger}.

Note that the Bluetooth specification defines an extended advertisement report~\cite[p. 2402]{bt52}.
However, this report type also does not contain channel information.

\begin{figure}[!t]
\begin{lstlisting}[caption={Enhanced event type interpretation.}, label=lst:event]
@\textcolor{blue!70!black}{channel}@   = (event_type >> 4) & 7
@\textcolor{blue!70!black}{antenna}@   = event_type & 0x80
@\textcolor{blue!70!black}{scan\_mode\hspace{0.2em}}@ = (event_type >> 3) & 3
\end{lstlisting}
\end{figure}

\subsection{Bluetooth Firmware}
\label{ssec:firmware}

The \emph{PacketLogger} reverse-engineering only indicates that additional event types exist. They still need to be enabled within
the firmware. For the firmware analysis, we dump firmware from a selection of chips with \emph{InternalBlue}.
\emph{WICED Studio 6.2} contains partial symbols for various \emph{Cypress} chips as well as the \emph{Broadcom} \emph{BCM20703A2} chip that is in 
\emph{MacBooks} produced in 2016 and 2017.

The global boolean \path{bEnhancedAdvReport} changes the behavior of the functions \path{_scanTaskRxHeaderDone} and \path{lculp_HandleScanReport}. This is explained in the next two paragraphs.

The firmware is organized in tasks that are called by the \ac{BCS}. The scan task is responsible for receiving advertisements.
In general, packet reception tasks are separated into receiving a header and receiving the according payload. 
The channel is already known when receiving a header, and thus, if \path{bEnhancedAdvReport} is set, additional information is copied
from the raw packet data into \path{ulp_extraInfo}.

While tasks need to be finished within the strict timings of the Bluetooth clock, handlers asynchronously parse task data and pass it on to the host's operating system driver.
In the case of an advertisement, this handler is \path{lculp_HandleScanReport}.
The prefix \path{lculp} stands for link control in the ultra-low-power protocol, namely \ac{BLE}. 
The advertisement handler copies the additional information into the \path{event_type} field if \path{bEnhancedAdvReport} is set.

Searching for the variable name \path{bEnhancedAdvReport} is only possible within a firmware that has partial symbols. However, symbols for most
off-the-shelf devices are unknown. Nonetheless, hardware registers are mapped similarly over various firmware generations. Also, the architecture is typically
an \emph{ARM Cortex M}, and compiler options are similar. Thus, we can search for equal \SI{4}{\byte} and \SI{8}{\byte} snippets,
which only return a few results within each firmware, to identify the scan task handler across multiple firmware versions.
In our case, the scan task handler disassembly includes the line \textcolor{gray}{\texttt{mov.w r0, \#0x650000}},
which is a \SI{4}{\byte} instruction represented by \textcolor{gray}{\texttt{0x00caf44f}}, and that we used
for manual binary diffing.

We find that the comparably old \emph{Nexus 5} firmware with a build date from 2012 does not feature the \path{bEnhancedAdvReport} flag.
However, the \emph{Cypress} evaluation boards \emph{CYW20719}, \emph{CYW20735}, and \emph{CYW20819} support it, as well as the \emph{MacBook} 2016--2017 chip \emph{BCM20703A2},
the \emph{MacBook} 2017--2019 chip \emph{BCM4364B0}, and the \emph{Samsung Galaxy S10/S20} chip \emph{BCM4375B1}. We assume that this feature was introduced
by \emph{Broadcom} around 2014 and all newer chips support it. While our \emph{InternalBlue}-based setup can only enable this feature for research,
\emph{Broadcom} could also roll it out as a patch for a broad variety of devices.

%% file: chapters/conclusion.tex

\section{Conclusion}
\label{sec:conclusion}

The channel reporting within \ac{BLE} advertisements can be enabled with a single flag
on most \emph{Broadcom} and \emph{Cypress} chips. This makes the patch rather simple, as it only needs to set the flag
to \path{0x01}.
Future practical tests within contact tracing will show how much channel awareness
can improve proximity measurements.
Further physical-layer properties might also be available during advertisement
reception. However, we did not spot any additional advertisement flags during the reverse-engineering process.
Thus, more complex patches are required for further physical-layer insights.

%% file: chapters/demo.tex

\section*{Demo}


This demo consists of an \emph{InternalBlue} addition that enables the enhanced event type within advertisements on various
\emph{Broadcom} and \emph{Cypress} chips. \emph{InternalBlue} runs on \emph{Android}, \emph{Linux}, \emph{macOS},  and \emph{iOS}.
Thus, the demo can be tested on various devices, as long as they have a supported chip.

\emph{InternalBlue} is open-source and available on \url{https://github.com/seemoo-lab/internalblue}. The extended advertisements
can be activated by running the command \texttt{adv}.
After enabling the enhanced advertisements with \texttt{adv} and opening \emph{Wireshark},
the event type field in advertisements contains the masked channel, antenna, and scan mode information.

To show this demo, we will provide a video of receiving the enhanced event type on a \emph{Cypress CYW20819} evaluation
board on \emph{Linux}. On the \emph{Linux BlueZ} stack, advertisements can be received by executing \path{hcitool} \path{lescan}.

\begin{figure}[h]
\includegraphics[width=1.0\columnwidth]{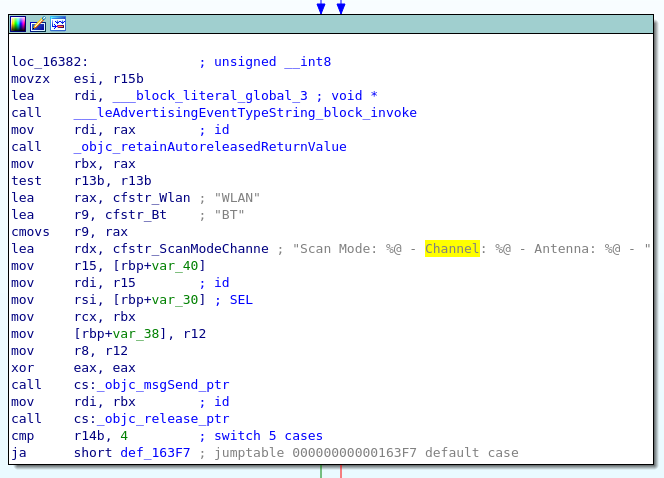}
\caption{\emph{PacketLogger} analysis in \emph{IDA Pro 7.2}.}
\label{fig:ida}
\end{figure}

Moreover, the video will show the workflow of reverse-engineer\-ing \emph{PacketLogger} to identify the field where the channel
information is included, as depicted in \autoref{fig:ida}. Then, we will search for this field in a partially symbolized \emph{Cypress} firmware. This part of the demo will help
other researchers to identify similar proprietary features.

\newpage